\documentclass[%
 reprint,
superscriptaddress,
 amsmath,amssymb,
 aps,
twocolumn,
]{revtex4-2}

\usepackage{xcolor}
\usepackage{graphicx}
\usepackage{dcolumn}
\usepackage{bm}
\usepackage[colorlinks,allcolors=blue,urlcolor=blue]{hyperref}

\newcommand{\hermconj}[1]{{#1}^\dagger}
\newcommand{\pseudoinv}[1]{{#1}^+}

\begin{document}

\preprint{APS/123-QED}

\title{Dynamic mode decomposition of nonequilibrium electron-phonon dynamics:\\ accelerating the first-principles real-time Boltzmann equation}

\author{Ivan Maliyov}
\affiliation{
 Department of Applied Physics and Materials Science, \protect\\ California Institute of Technology, Pasadena, California 91125
}

\author{Jia Yin}
\affiliation{Applied Mathematics and Computational Research Division, Lawrence Berkeley National Laboratory, Berkeley, California, 94720}

\author{Jia Yao}
\affiliation{
 Department of Applied Physics and Materials Science, \protect\\ California Institute of Technology, Pasadena, California 91125
}
\affiliation{
 Department of Physics, California Institute of Technology, Pasadena, California 91125
}

\author{Chao Yang}
\affiliation{Applied Mathematics and Computational Research Division, Lawrence Berkeley National Laboratory, Berkeley, California, 94720}

\author{Marco Bernardi}%
\affiliation{
 Department of Applied Physics and Materials Science, \protect\\ California Institute of Technology, Pasadena, California 91125
}
\affiliation{
 Department of Physics, California Institute of Technology, Pasadena, California 91125
}


\begin{abstract}

Nonequilibrium dynamics governed by electron-phonon ($e$-ph) interactions plays a key role in electronic devices and spectroscopies and is central to understanding electronic excitations in materials.
The real-time Boltzmann transport equation (rt-BTE) with collision processes computed from first principles can describe the coupled dynamics of electrons and atomic vibrations (phonons). 
Yet, a bottleneck of these simulations is the calculation of $e$-ph scattering integrals on dense momentum grids at each time step. 
Here we show a data-driven approach based on dynamic mode decomposition (DMD) that can accelerate the time propagation of the rt-BTE and identify dominant electronic processes. 
We apply this approach to two case studies, high-field charge transport and ultrafast excited electron relaxation.
In both cases, simulating only a short time window of $\sim$10\% of the dynamics suffices to predict the dynamics from initial excitation to steady state using DMD extrapolation. 
Analysis of the momentum-space modes extracted from DMD sheds light on the microscopic mechanisms governing electron relaxation to steady state or equilibrium.
The combination of accuracy and efficiency makes our DMD-based method a valuable tool for investigating ultrafast dynamics in a wide range of materials. 
\end{abstract}

\maketitle

\section{Introduction}
\vspace{-10pt}
First-principles calculations are widely employed for modeling and designing materials, with applications ranging from energy~\cite{Montoya_2016,Pham_2017} to (opto)electronic devices~\cite{Jungwirth_2006,Charlier_2007,Cuevas_2017} to materials discovery ~\cite{Curtarolo_2013,Jain_2016,Marzari_2021}. 
Starting from the crystal structure and atomic positions as the main inputs, these methods can predict material properties including mechanical, electrical, magnetic and optical. 
While computing ground-state and linear-response properties with density functional theory (DFT) is a decades-long effort~\cite{Martin,Reining,Burke_2012,Baroni_2001}, recent work has focused on modeling ultrafast dynamics in materials and simulating time-domain spectroscopies from first-principles~\cite{bernardi-si, Sangalli2015, Jhalani2017, Sjakste, tong2020precise, Caruso_2021, Calandra-1, Prezhdo2021, Perfetto2022, Zheng_2023, Bernardi_2023}. These methods focused on nonequilibrium dynamics are a more recent research frontier with both theoretical and computational challenges.

First-principles calculations in the time domain provide a microscopic description of nonequilibrium dynamics in materials. These methods propagate in time quantities characterizing the quantum dynamics, such as the time-dependent electron wave function, density~\cite{marques2012fundamentals}, density matrix~\cite{Molina2017} or Green's function~\cite{Perfetto2022}, and can also access the time-dependent atomic positions and lattice vibrations~\cite{tong2020precise, Caruso_2021}.
Different schemes are successful in different regimes. For example, coherent electron dynamics on the attosecond time scale can be modeled effectively using time-dependent DFT~\cite{Rozzi2013, Maitra_2016, Atto_nature, Lloyd_Hughes_2021, Zheng_2023}, but that approach is not ideal for
modeling phonon dynamics, which occurs on a picosecond time scale~\cite{Kononov_2022}.
\\
\indent
The real-time Boltzmann transport equation (rt-BTE) has emerged as an effective tool for exploring the coupled electron and phonon dynamics from femtosecond to nanosecond timescales~\cite{Jhalani2017,Maliyov_2021, Caruso_2021}. 
In the rt-BTE, the time-dependent electron populations are obtained by solving a set of integro-differential equations accounting for $e$-ph scattering processes on dense momentum grids. Following an initial excitation, the rt-BTE is propagated in time to reach thermal equilibrium or steady state in an external field. 
This scheme employs a femtosecond time step to capture the $e$-ph scattering processes. However, evaluating the scattering integral at each time step makes the rt-BTE approach computationally demanding even for materials with a handful of atoms in the unit cell.
\\
\indent
Data-driven techniques are increasingly employed in materials modeling, both for accelerating computational workflows and 
to gain physical insight using learning \mbox{algorithms~\cite{Ramprasad_2017,Schmidt_2019}}.
In particular, dynamic mode decomposition (DMD), which was developed in the last decade to study fluid dynamics, is a valuable tool to linearize dynamical problems and reduce their dimensionality~\cite{SCHMID_2010,kutz2016dynamic}. 
In DMD, explicit simulation of a short initial time window allows one to learn the dominant modes governing the dynamics and extrapolate the simulation to future times at low computational cost.
Recent work has employed DMD to study electron dynamics described by model Hamiltonians with purely electronic interactions~\cite{Yin_2023,Reeves_2023}. Yet, to date DMD has not  been applied to more computationally intensive first-principles studies.
\\
\indent
In this work, we combine DMD with first-principles calculations of nonequilibrium electron dynamics, using the framework of the rt-BTE in the presence of $e$-ph collisions and external fields. 
We show that DMD provides an order-of-magnitude computational speed-up while retaining the full accuracy of the first-principles rt-BTE. In addition, DMD reveals key momentum-space temporal patterns and achieves a significant dimensionality reduction of the nonequilibrium physics. 
Our results include both high-field transport and transient excited-state dynamics, and are accompanied by a careful characterization of convergence with respect to the size of the sampling window during which DMD learns the dominant modes. 
Taken together, this work provides the blueprint for combining data-driven methods with first-principles calculations to study nonequilibrium dynamics in real materials.

\begin{figure*}
\centering
\includegraphics[width=0.95\textwidth]{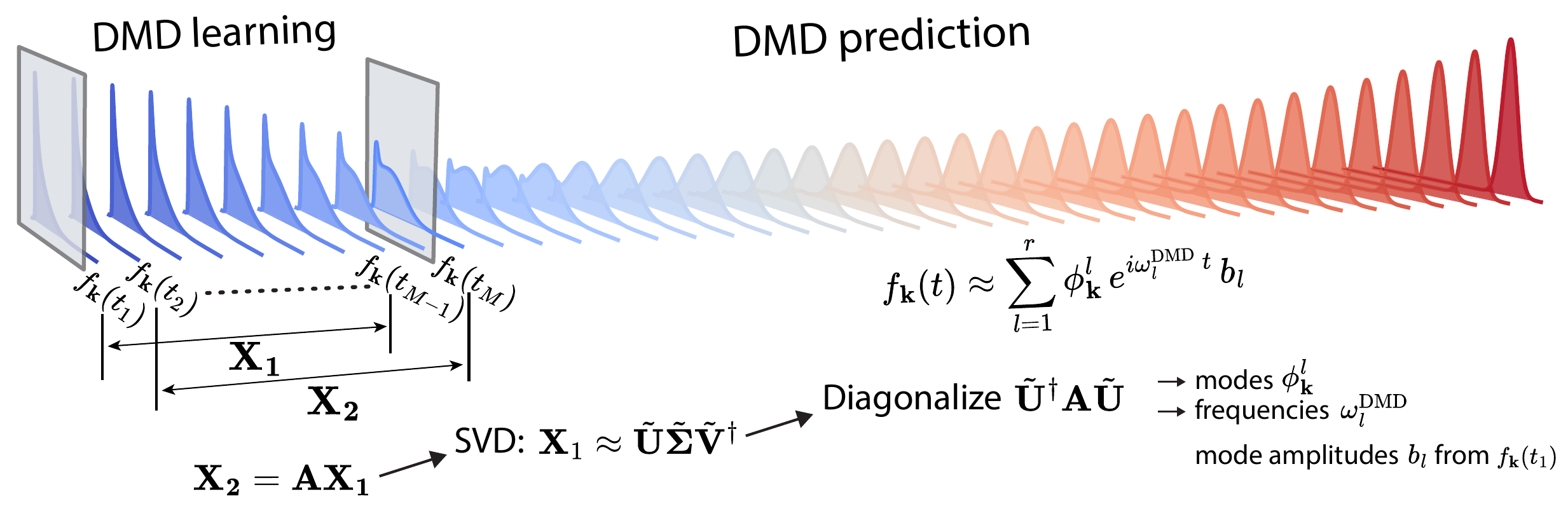}
\caption{\textbf{Workflow of DMD plus rt-BTE calculations.} The first $M$ steps of the dynamics, which make up the sampling window for DMD learning, are simulated by solving the rt-BTE. The resulting populations $f_{\mathbf{k}}(t)$ are stacked in the $\mathbf{X}_1$ and $\mathbf{X}_2$ matrices with a relative shift of one time step. The dynamics at later times $t > t_M$ is predicted with DMD using the $r$ leading modes obtained by SVD of the matrix $\mathbf{X}_1$ and diagonalization of the matrix $\tilde{\mathbf{A}} = \hermconj{\tilde{\mathbf{U}}} \mathbf{A} \tilde{\mathbf{U}}$.}  
\label{fig1}
\end{figure*}

\section{Results}

\subsection{First-principles rt-BTE}
\vspace{-10pt}
We describe the electron distribution using the time-dependent populations $f_{n\mathbf{k}}(t)$, which quantify the occupation of each electronic state $|n\mathbf{k}\rangle$, where $\mathbf{k}$ is the electron crystal momentum and $n$ is the band index (from now on we omit the band index to simplify the notation).
Starting from an initial distribution at time zero, $f_{\mathbf{k}}(t=0)$, in the rt-BTE the populations evolve according to~\cite{zhou2021perturbo}
\begin{equation}
\label{eq:rt-bte}    
\frac{\partial f_{\mathbf{k}}(t)}{\partial t} = \frac{\mathbf{F}}{\hbar}\cdot \nabla _\mathbf{k}f_{\mathbf{k}}(t)+\mathcal{I}[f_{\mathbf{k}}(t)],
\end{equation}
where $\mathcal{I}[f_{\mathbf{k}}(t)]$ is the collision integral accounting for $e$-ph scattering processes in momentum space \cite{mahan2010condensed} and $\mathbf{F}$ includes any external fields applied to the system. 
\\
\indent
The rt-BTE simulations use dense momentum grids to accurately describe scattering between electronic states via absorption and emission of phonons. The required grid sizes are typically greater than $100 \times 100 \times 100$ for both electron and phonon momenta. We time-step equation \ref{eq:rt-bte} using  explicit solvers (Euler or 4th-order Runge-Kutta) or more advanced Strang splitting techniques~\cite{Maliyov_2021}. 
The collision integral includes a summation over the phonon momentum grid and is evaluated at least once per time step using a parallel algorithm implemented in the {\sc Perturbo} code~\cite{zhou2021perturbo} (see Methods for details). 
Although here we focus on the dynamics of electrons interacting with phonons, the \mbox{rt-BTE} formalism has also been extended to study nonequilibrium phonon~\cite{tong2020precise} and exciton dynamics~\cite{Chen-excitons}.
\subsection{\label{subseq:DMD-scheme} DMD learning and prediction of the dynamics}
\vspace{-10pt}
We employ DMD in combination with rt-BTE simulations. 
The DMD approach linearizes the dynamics by relating the states of the system at times $t$ and $t + \Delta t$ via a time-independent matrix $\mathbf{A}$~\cite{Koopman_1931,ROWLEY_2009}. 
Focusing on the $e$-ph dynamics, this amounts to advancing the electronic populations at time $t$ using
\begin{equation}
\label{eq:f2Af1}    
f_{\mathbf{k}}(t + \Delta t) = \mathbf{A} f_{\mathbf{k}}(t),
\end{equation}
where the populations $f_{\mathbf{k}}$ form a vector with size $N$ equal to the number of $\mathbf{k}$-points in the electronic momentum grid (typically, $N\approx 10^5 - 10^6$).
To obtain the matrix $\mathbf{A}$, we time-step the rt-BTE in a sampling window consisting of $M$ time steps (using the {\sc Perturbo} code~\cite{zhou2021perturbo}) and then we form two matrices $\mathbf{X}_1$ and $\mathbf{X}_2$ relating the populations at times $t$ and $t+\Delta t$.  
The populations $f_{\mathbf{k}}(t)$ from $t_1$ to $t_{M-1}$ are stacked column-wise in the matrix $\mathbf{X}_1$, with column $i$ corresponding to time $t_i$ and containing the populations $f_{\mathbf{k}}(t_i)$ for all $\mathbf{k}$-points and bands. The populations from $t_2$ to $t_M$ are similarly stacked column-wise in the second matrix $\mathbf{X}_2$. 
\\
\indent
According to equation \ref{eq:f2Af1}, these matrices are related by $\mathbf{X}_2 = \mathbf{A} \mathbf{X}_1$, but 
computing $\mathbf{A}$ naively from the pseudoinverse of $\mathbf{X}_1$ has a prohobitive cost due to the large size $N$ of the $\mathbf{k}$-point grid. 
To circumvent this problem, in DMD one first performs a truncated singular value decomposition (SVD)~\cite{Golub_1965,golub2013matrix} of the $\mathbf{X}_1$ matrix: 
\begin{equation}
\label{eq:svd1}
    \mathbf{X}_1 = \mathbf{U} \mathbf{\Sigma} \hermconj{\mathbf{V}},
\end{equation}
where $\mathbf{\Sigma} \in \mathbb{R}^{N\times M}$ is a matrix with diagonal entries equal to the singular values $\sigma_j$ arranged in decreasing order, while $\mathbf{U} \in \mathbb{C}^{N\times N}$ and $\mathbf{V} \in \mathbb{C}^{M\times M}$ are matrices collecting the mutually orthogonal singular vectors~\cite{brunton2022data}. 
(Above, $\hermconj{\mathbf{V}}$ indicates the Hermitian conjugate of $\mathbf{V}$.) 
\\
\indent
Because $\mathbf{X}_1$ contains the time-dependent populations, this SVD procedure can single out the main patterns in the momentum-space dynamics. Here, we keep only the first $r$ singular values (typically, $r\approx 10$) to restrict the solution space to the leading $r$ momentum-space modes, and then project the matrix $\mathbf{A}$ onto this reduced $r$-dimensional space. This procedure provides the matrix $\mathbf{\tilde{A}}$, with reduced size $r \times r$, which can be diagonalized straightforwardly to obtain the dominant DMD modes.
Using this procedure, the populations at future times $t > t_M$ are \textit{predicted} $-$ that is, obtained without explicit solution of the rt-BTE $-$ using
\begin{equation}
\label{eq:DMD-form}    
f_{\mathbf{k}}(t>t_M) \approx \sum_{l=1}^r \, b_l\,\phi^l_{\mathbf{k}} \, e^{ i \omega_l^{\mathrm{DMD}}\, t},
\end{equation} 
where $\phi^l_{\mathbf{k}}$ are the momentum-space DMD modes obtained from the matrix $\mathbf{\tilde{A}}$, and $\omega_l^{\mathrm{DMD}}$ and $b_l$ are their frequencies and amplitudes (see Methods for detailed derivations). 
\\
\indent
We summarize the main steps of this DMD procedure, which are illustrated in Fig. \ref{fig1}:

\begin{enumerate}
    \item Simulate the rt-BTE dynamics for the first $M$ steps and construct the matrices $\mathbf{X}_1$ and $\mathbf{X}_2$;
    \item Perform SVD on $\mathbf{X}_1$ to find the matrix $\mathbf{\tilde{A}}$ in the reduced $r$-dimensional space;
    \item Diagonalize $\mathbf{\tilde{A}}$ to find the DMD modes $\phi^l_{\mathbf{k}}$ and their frequencies $\omega_l^{\mathrm{DMD}}$, with $l=1\ldots r$;
    \item Obtain the mode amplitudes $b_l$ from the initial condition $f_{\mathbf{k}}(t_1)$;
    \item Predict the dynamics for $t > t_M$ using equation \ref{eq:DMD-form}.
\end{enumerate}
A key parameter is the duration of the sampling window ($t_M$) required for accurate DMD extrapolation of the dynamics beyond $t_M$. As the computational cost of DMD is negligible, the size of the sampling window, during which the rt-BTE is solved by explicit time-stepping, determines the computational cost of the entire workflow. 
\subsection{High-field electron dynamics}
\vspace{-15pt}
\begin{figure*}[t]
\centering
\includegraphics[width=2\columnwidth]{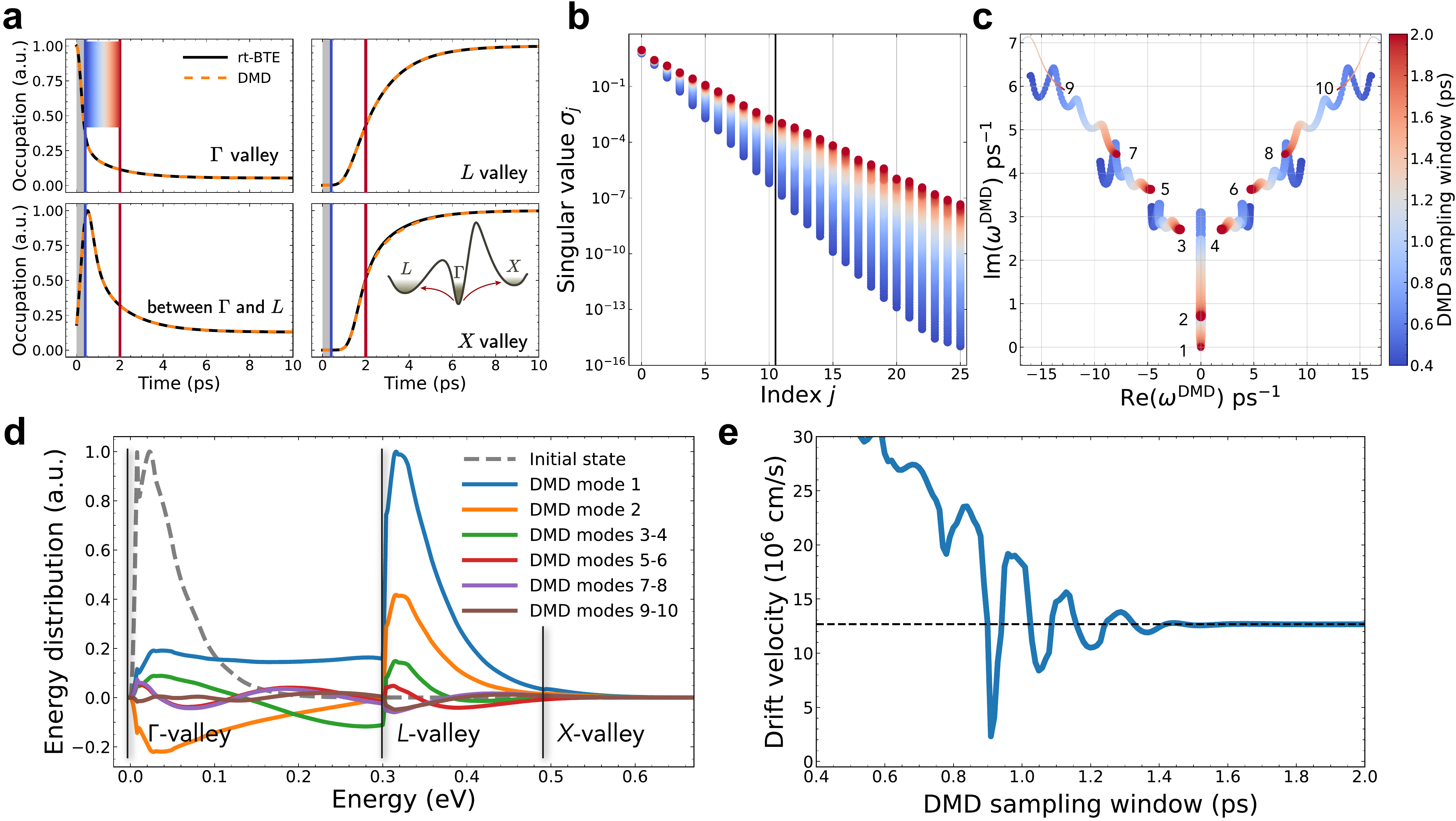}
\caption{\textbf{DMD simulations of electrons in GaAs in an applied electric field}. 
\textbf{a}, Time-dependent electron populations for four electron momenta. The gray region indicates the shortest sampling window (0.4 ps) and the red vertical line the longest sampling window we tested (2 ps). Solid black lines show the rt-BTE results and orange dashed lines the DMD predictions obtained using the longest sampling window. The inset is a schematic of the low-energy band structure of GaAs showing the $\Gamma$- and higher energy $L$- and $X$-valleys.
\textbf{b}, Singular values of the $\mathbf{X}_1$ matrix, with the ten largest singular values used in our DMD calculations separated by a vertical line.
\textbf{c}, DMD frequencies plotted in the complex plane and shown as circles with radii proportional to the DMD mode amplitudes $b_l$. In panels \textbf{a}, \textbf{b}, and \textbf{c}, the colors indicate the duration of the DMD sampling window according to the legend given in (\textbf{c}). 
\textbf{d}, DMD momentum-space modes $\phi^l_{\mathbf{k}}$, multiplied by the corresponding amplitudes $b_l$, given as a function of energy. The initial state $f_\mathbf{k}(t_1)$ is shown with a dashed line. 
\textbf{e}, Convergence of the steady-state drift velocity with respect to duration of the DMD sampling window. The rt-BTE value is shown for reference as a dashed line.
} 
\label{fig2}
\end{figure*}
We employ our DMD-based approach to simulate time-domain electron dynamics in an applied electric field in the presence of $e$-ph collisions.
We recently demonstrated similar calculations using the rt-BTE without the aid of data-driven techniques~\cite{Maliyov_2021}.
Here we use this case study to explore the accuracy and efficiency of our rt-BTE plus DMD approach as well as find optimal values for the sampling window and analyze the momentum-space DMD modes.
%
Our calculations focus on electrons in GaAs, where the  conduction band has three sets of low-energy valleys, at $\Gamma$ and near $L$ and $X$ in order of increasing energy~\cite{yu-cardona} (see the inset in Fig.~\ref{fig2}a). 
Upon applying an electric field, the electrons are accelerated to higher band energies while they also transfer part of that excess energy to the lattice via $e$-ph collisions. These competing mechanisms lead to a steady-state electronic distribution which is typically reached on a picosecond to nanosecond time scale.  
\\
\indent
Our simulations begin with electrons in thermal equilibrium with the lattice at 300~K. We apply a constant electric field $\mathbf{E}$ and time step the electron populations until they reach the steady state distribution, $f_{\mathbf{k}}^\mathbf{E}$, from which we compute the mean drift velocity, $v(\mathbf{E})$, a quantity routinely measured in experiments~\cite{canali1971drift,ashida1974energy,ferry1975high}. Repeating this procedure for multiple field values allows us to construct the full drift velocity versus electric field curve in a material, starting from linear response at low field to velocity saturation at high field~\cite{Maliyov_2021}.
\\
\indent
Figure~\ref{fig2}a shows the time-dependent populations in four regions of the Brillouin zone following the application of a high field (5 kVcm$^{-1}$). Electrons initially occupying the $\Gamma$-valley scatter to the higher-energy $L$- and $X$-valleys. As a result, the electron populations in the $\Gamma$-valley decrease, with a corresponding increase in $L$- and $X$-valley populations. 
In regions of momentum space between the $\Gamma$- and $L$-valleys the populations peak at intermediate times and then relax to lower \mbox{values.}
\\
\indent
This dynamics is nontrivial because the populations evolve differently in different momentum-space regions, making accurate predictions challenging.
Our DMD approach can learn the dominant modes governing this intricate dynamics and extrapolate the time-dependent populations well beyond the sampling window. Remarkably, we find that a short sampling \mbox{window} $-$ 400~fs to 2~ps out of a total simulation time of 12.5~ps $-$ is sufficient to extrapolate the dynamics all the way to steady state, with rt-BTE and DMD trajectories in nearly exact agreement outside the sampling window~(Fig.~\ref{fig2}a). 
This accuracy extends to the entire set of $\sim$10$^5$ $\mathbf{k}$-points considered in our simulations. 
\\
\indent
The ability to learn key temporal momentum-space patterns is a consequence of the relatively rapid decay of the singular values of the $\mathbf{X}_1$ matrix used for learning the dynamics in the sampling window (Fig.~\ref{fig2}b). This decay becomes slower as the sampling window increases, but it remains significant even for the longest sampling window of 2~ps used here (note the log scale in the plot). 
In turn, the singular value decay enables a striking dimensionality reduction, with DMD employing only $r \!\approx \!10$ modes to solve the dynamics as opposed to $10^5$ populations $f_{\mathbf{k}}$ and billions of $e$-ph scattering terms in the rt-BTE. 
\\ 
\indent
The choice of an ideal sampling window can rely on the appearance of specific DMD modes at steady state.  
Figure~\ref{fig2}c shows the DMD mode frequencies $\omega^{\mathrm{DMD}}$ in the complex plane, where the imaginary part of $\omega^{\mathrm{DMD}}$ corresponds to the decay rate of a given mode and the real part gives its oscillation frequency. 
The populations $f_{\mathbf{k}}(t)$ are real-valued and are written as a summation of complex exponentials in equation~\ref{eq:DMD-form}. Therefore, physically meaningful results are possible only when $\mathrm{Re}(\omega^{\mathrm{DMD}})=0$ (modes 1, 2) or when $\omega^{\mathrm{DMD}}$ appear as complex conjugate pairs (modes 3 $-$ 10).
Describing the steady state is particularly important in our simulations. In DMD, all modes with a non-zero imaginary frequency vanish in the long time limit, with only one mode surviving at steady state (mode 1 in Fig.~\ref{fig2}c). 
As the sampling window increases, the imaginary frequency of this mode goes to zero, providing the correct steady state behavior. This analysis allows us to find the minimal sampling window required for accurate steady-state results by monitoring the zero-frequency mode.
\\
\indent 
The DMD eigenvector of the zero-frequency mode (mode 1 in Fig.~\ref{fig2}d) 
determines the steady-state electron distribution $\phi^1_{\mathbf{k}} = f_{\mathbf{k}}(t\rightarrow \infty)$, while the other modes control the transient dynamics.
For example, mode 2 governs electron scattering from the $\Gamma$- to the $L$- and $X$-valleys, and higher modes appearing as conjugate pairs exhibit oscillating trends in energy (modes 3$-$10 in Fig.~\ref{fig2}d).
Converging the zero-frequency mode allows us to compute the steady-state drift velocity more efficiently. 
Figure~\ref{fig2}e shows that a sampling window of 1.7~ps (170 snapshots) provides a drift velocity nearly identical to the full rt-BTE calculation, which requires much longer simulation times of up to 12.5~ps (1250 snapshots). 
On this basis we conclude that DMD needs only $\sim$10\% of the dynamics data for accurate steady-state predictions.
%

\subsection{\label{subsec:vel-field}
Velocity-field curves}
\vspace{-10pt}
\begin{figure}[t]
\centering
\includegraphics[width=1.0\columnwidth]{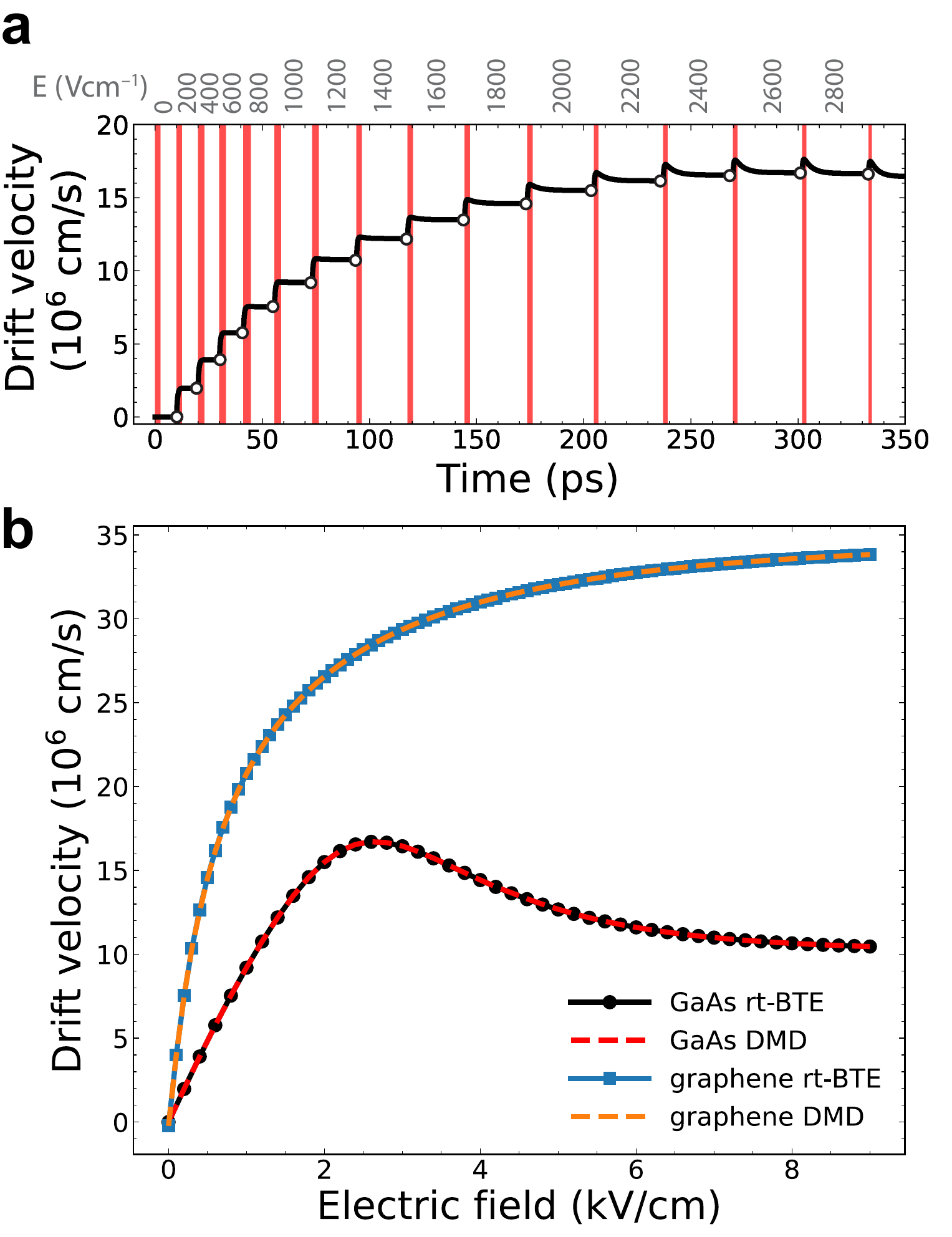}
\caption{\textbf{Velocity-field curves from DMD.}
\textbf{a}, Transient drift velocity in GaAs computed as a function of time (black curve). The external electric field is increased step-wise in the simulation (see the field values given above the plot). The DMD sampling window for each electric field is shown with a red rectangle, and the drift velocities outside this window are predicted with DMD. The steady-state drift velocities from DMD correspond to the plateaus for each field value, and agree with the reference drift velocities, obtained by explicitly time-stepping the rt-BTE until steady state, which are shown with white dots.
\textbf{b}, Velocity-field curves in GaAs and graphene obtained from the rt-BTE (black and blue solid lines) and by combining the rt-BTE and DMD (red and orange dashed lines).}
\label{fig3}
\end{figure}

We also employ DMD to accelerate calculations of entire velocity-field curves. This requires the drift velocity for a set of electric field values, and thus we adopt a modified workflow. 
Following our recent work~\cite{Maliyov_2021}, we gradually increase the electric field (black curve in Fig.~\ref{fig3}a) and use the steady-state populations for a given field, $f_{\mathbf{k}}^{\mathbf{E}}$, as the initial condition for the next field value, $E+\Delta E$, where the field increment $\Delta E$ is typically $100-200$~Vcm$^{-1}$. As the applied field increases, the DMD frequencies and momentum-space modes change substantially. 
Therefore, for each new field value we repeat DMD learning in the initial stage of the simulation (see the DMD sampling regions shown as red rectangles in Fig.~\ref{fig3}a). We then use DMD to predict the steady-state populations $f^{\mathbf{E}}_{\mathbf{k}}$ and drift velocity for that field value, and use $f^{\mathbf{E}}_{\mathbf{k}}$ as the initial condition for the next field value. 
\\
\indent
The velocity-field curves obtained with this approach are shown in Fig.~\ref{fig3}b for GaAs and graphene and compared with rt-BTE results obtained without DMD. 
Using DMD lowers significantly the computational cost to obtain the full velocity-field curves, by a factor of 10.5 for GaAs and $\sim$16.5 for graphene, while fully preserving the accuracy.
This efficiency is a consequence of DMD's ability to capture the dominant modes in the population dynamics using only a small number of snapshots, with a similar accuracy regardless of the electric field value. 
Our strategy of gradually increasing the electric field leads to an easier-to-extrapolate dynamics compared to the abrupt application of a strong field. With our approach, the DMD modes exhibit smooth trends in momentum space and the DMD frequencies are relatively small in magnitude, thus enabling accurate DMD predictions.\\

\vspace{0.2cm}
\subsection{\label{sec:cooling} Excited electron relaxation}
\begin{figure}[t]
\centering
\includegraphics[width=1.0\columnwidth]{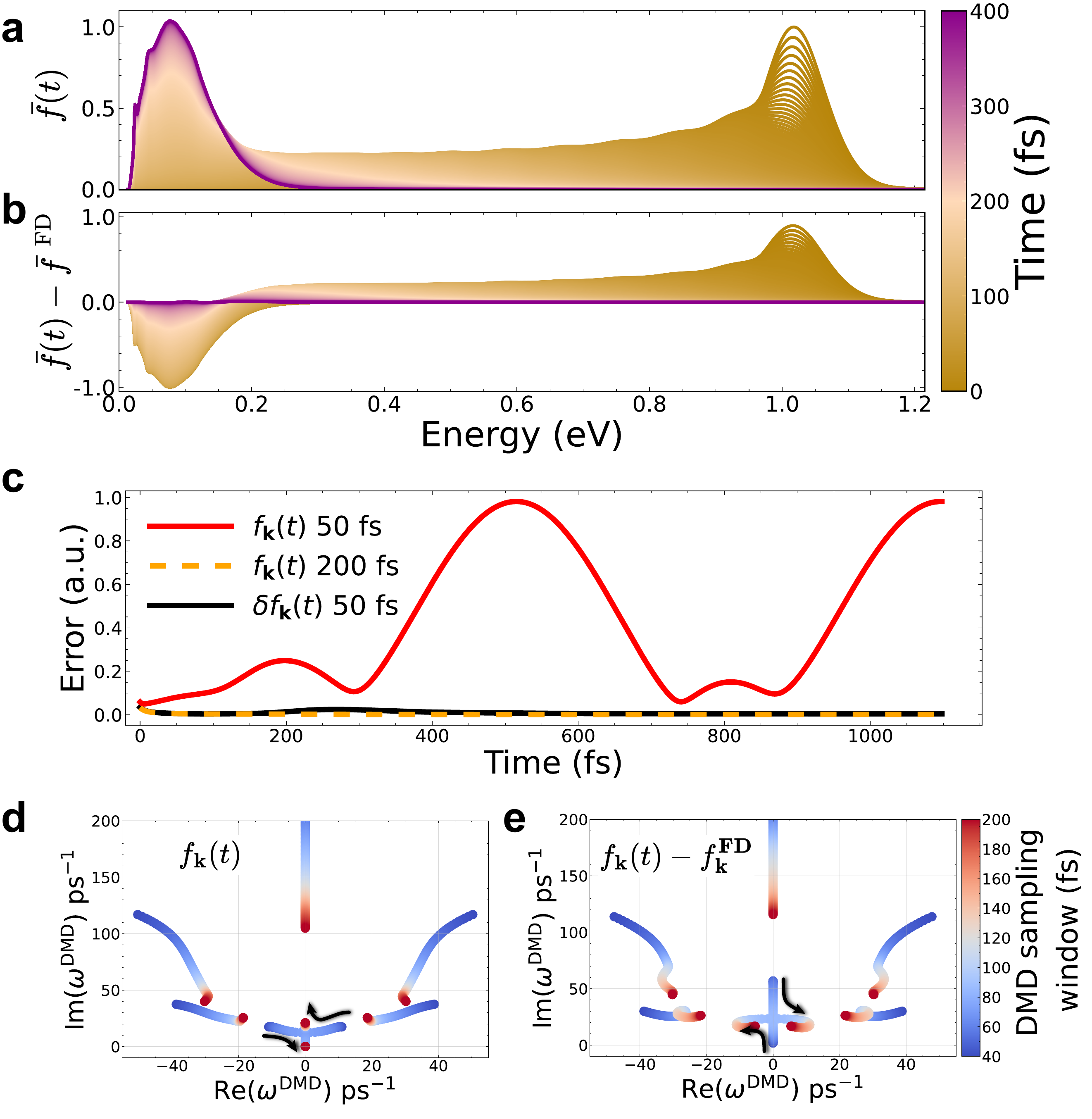}
\caption{\textbf{Transient dynamics in GaN using DMD.}
\mbox{\textbf{a}, Energy} dependence of the momentum-averaged electron populations, and \textbf{b}, difference between the time-dependent and equilibrium populations in GaN. In both panels, the simulation time is color-coded using sepia for the initial excited state and purple for the equilibrium state. The energy zero is set to the conduction band minimum. 
\textbf{c}, DMD error on the electron populations, computed as the root-mean-square difference between the reference values from rt-BTE and those obtained from DMD. 
Results are shown for different sampling windows, given in the legend, and for the two schemes where DMD is applied to $f_{\mathbf{k}}(t)$ or alternatively to $\delta f_{\mathbf{k}}(t) = f_{\mathbf{k}}(t)-f^{\mathrm{FD}}_{\mathbf{k}}$.
\textbf{d}, \textbf{e}, DMD frequencies on the complex plane, respectively for $f_{\mathbf{k}}(t)$ and $f_{\mathbf{k}}(t)-f^{\mathrm{FD}}_{\mathbf{k}}$, with  the duration of the sampling window color-coded.
\vspace{-10pt}} 
\label{fig4}
\end{figure}

\vspace{-10pt}
Next, we consider a different nonequilibrium dynamics where the material is initially prepared in an excited electronic state. This setting can be used, for example, to model the effect of an optical excitation with a laser pulse~\cite{bernardi-si}. Different from the high-field dynamics, in this case the long-time limit is known and we are primarily interested in the \textit{transient} dynamics. 
Following the initial excitation, in the presence of $e$-ph interactions and without any external fields, the electrons relax to a thermal equilibrium Fermi-Dirac distribution~\cite{ferry_book}, $f_{\mathbf{k}}^{\mathrm{FD}}$, typically on a sub-picosecond time scale. This ultrafast dynamics can be modeled by time-stepping the rt-BTE until reaching the equilibrium Fermi-Dirac distribution. 
Using this approach, our previous work has shown that electrons relax to the band edge significantly slower than holes in GaN semiconductor, with implications for optoelectronic devices~\cite{Jhalani2017}.
\\
\indent
Following that work, we model an excited state in GaN by placing the electrons $\sim$1~eV above the conduction band edge, and then obtain the time-dependent electron populations by solving the rt-BTE (see Fig.~\ref{fig4}a,b).
\mbox{We employ} DMD to predict this transient dynamics, and find large errors when using a short time window of up to $\sim$50 fs (solid red line in Fig.~\ref{fig4}c). The correct steady state and transient dynamics are obtained by increasing the sampling window to 200~fs (dashed orange line in Fig.~\ref{fig4}c). 
Our analysis of the DMD frequencies shows that the zero-frequency mode describing thermal equilibrium in the long-time limit appears when the sampling window reaches 100~fs (see the arrow in Fig.~\ref{fig4}d) and fully converges for a $\sim$200~fs sampling window. 
The need for such a long sampling window relative to the total duration of the dynamics (400 fs) makes DMD ineffective.
\\
\indent
To address this issue and more efficiently study transient dynamics with DMD, we formulate a different learning procedure that incorporates knowledge of the equilibrium state. 
We focus on the difference between the transient and equilibrium populations, $\delta f_{\mathbf{k}}(t) = f_{\mathbf{k}}(t) - f^{\mathrm{FD}}_{\mathbf{k}}$, as opposed to just $f_{\mathbf{k}}(t)$ as we did in the high-field example. After predicting $\delta f_{\mathbf{k}}$(t) with DMD, we obtain the time-dependent populations $f_{\mathbf{k}}(t)$ by adding back the $f^{\mathrm{FD}}_{\mathbf{k}}$ term. 
As $\delta f_{\mathbf{k}}$ vanishes in the long-time limit (Fig.~\ref{fig4}b), the zero-frequency DMD mode is missing when computing $\delta f_{\mathbf{k}}$ (Fig.~\ref{fig4}e);  
all other DMD frequencies associated with $\delta f_{\mathbf{k}}$ are similar to those for $f_{\mathbf{k}}(t)$ (Fig.~\ref{fig4}d-e). We find that the DMD method based on $\delta f_{\mathbf{k}}$ is far more effective and requires a significantly shorter sampling window for accurate DMD predictions $-$ 
using a 50~fs sampling window, we achieve results similar to DMD for $f_{\mathbf{k}}(t)$ with a four times longer (200~fs) window (Fig.~\ref{fig4}c).

With this improved DMD approach, using a sampling window of only $\sim$12\% of the total simulation time allows us to accurately predict the average electron relaxation rate in GaN, with a DMD computed value of 5.23~eVfs$^{-1}$ in close agreement (within 0.8\%) with the rt-BTE result. This result demonstrates that our DMD approach can predict excited electron relaxation with a high accuracy.

\section{Discussion}
\vspace{-10pt}
The DMD approach introduced here is highly efficient $-$ different from the rt-BTE, which requires a supercomputer with extensive memory and CPU resources, our DMD workflow can be performed straightforwardly on a laptop. 
The most demanding step is carrying out truncated SVD on the $\mathbf{X}_1$ matrix, but for comparison this step requires lower computational resources than even just a single rt-BTE time step. 
\\
\indent
This remarkable speed-up is achieved by reducing the dimensionality of the rt-BTE dynamics and is linked to the shape of the $\mathbf{X}_1$ matrix. The rt-BTE employs a large number of $\mathbf{k}$-points (about $10^5 - 10^6$), which equals the number of rows of the matrix $\mathbf{X}_1$, and a significantly smaller number of snapshots in the DMD sampling window, typically $\sim$100 time steps, which sets the number of columns in $\mathbf{X}_1$. 
Following truncated SVD, the size of the problem is reduced to (at most) the number of snapshots and is typically of order 50$-$100, and thus smaller by orders of magnitude compared to the original rt-BTE. (Note that one could use the entire set of singular values, but here we prefer using only $\sim$10 singular values to prevent numerical instabilities~\cite{brunton2022data}).
\\
\indent
This efficiency allows us to evaluate the accuracy of DMD on the fly, halting explicit time-stepping of the rt-BTE when the DMD steady state or transient dynamics are fully converged.
In addition, our approach addresses the key challenge of storing the rt-BTE populations, which are needed only in the sampling window in DMD, as opposed to the full dynamics. This is a critical improvement because in conventional rt-BTE simulations one needs to store the populations $f_{\mathbf{k}}(t)$ on dense momentum grids for thousands of time steps, resulting in terabytes of data. In contrast, after carrying out SVD in the sampling window, DMD stores only a handful of complex frequencies and momentum-space modes, using which the dynamics can be reconstructed for the entire simulation.\\

\section{Conclusion}
\vspace{-10pt}
In summary, we have introduced a data-driven approach based on DMD to accelerate first-principles calculations of nonequilibrium electron dynamics in \mbox{materials.} 
Our method speeds-up the solution of the time-dependent Boltzmann equation with electron collisions computed from first principles. 
We have shown that DMD can capture dominant modes governing the microscopic dynamics, enabling accurate predictions of the steady-state properties such as the drift velocity as well as transient processes such as electron relaxation and equilibration. 
In both steady-state and transient nonequilibrium calculations, DMD requires explicit time-stepping of the rt-BTE in a time window of only $\sim$10\% of the full simulation, after which the dynamics is extrapolated from the DMD modes with negligible computational cost.
This DMD workflow preserves the accuracy while requiring far more modest computational resources than full rt-BTE simulations.
\\
\indent
These advances are broadly \mbox{relevant} to studying nonequilibrium quantum dynamics of elementary \mbox{excitations.} For example, in future work our data-driven approach will be adapted to study phonon dynamics, which requires costly computations of phonon-phonon scattering, as well as exciton dynamics using a bosonic rt-BTE formalism.
\newpage
\section{Methods}

\subsection{Computing DMD modes and frequencies}
\vspace{-10pt}
Let us describe in more detail the calculation of DMD modes and frequencies. 
We start from the snapshots $f_{\mathbf{k}}(t)$ evaluated explicitly with the rt-BTE in the sampling window $t_1 < t < t_M$, and then apply the SVD procedure to the matrix $\mathbf{X}_1$ (see equation \ref{eq:svd1}). As shown in Fig.~\ref{fig2}b, we find that the singular values $\sigma_j$ decay rapidly. Keeping only the largest $r \approx 10$ singular values, we write the SVD of $\mathbf{X}_1$ as
\begin{equation}
    \mathbf{X}_1 \approx \tilde{\mathbf{U}} \tilde{\mathbf{\Sigma}} \hermconj{\tilde{\mathbf{V}}},
\end{equation}
where we defined the economy-sized matrices in the $r$-dimensional subspace~\cite{brunton2022data} as $\tilde{\mathbf{\Sigma}} = \mathbf{\Sigma}(1:r, 1:r)$, $\tilde{\mathbf{U}} = \mathbf{U}(1:N, 1:r)$, $\tilde{\mathbf{V}} = \mathbf{V}(1:M, 1:r)$. 
This way, the approximate pseudo-inverse of the matrix $\mathbf{X}_1$, denoted as $\pseudoinv{\mathbf{X}_1}$, can be obtained with little effort as $\tilde{\mathbf{V}} \tilde{\mathbf{\Sigma}}^{-1} \hermconj{\tilde{\mathbf{U}}}$. Then the matrix $\mathbf{A}$ relating the snapshot matrices via $\mathbf{X}_2 = \mathbf{A} \mathbf{X}_1$ can be written as
\begin{equation}
    \mathbf{A} = \mathbf{X}_2 \pseudoinv{\mathbf{X}_1} = \mathbf{X}_2 \tilde{\mathbf{V}} \tilde{\mathbf{\Sigma}}^{-1} \hermconj{\tilde{\mathbf{U}}}.
\end{equation}
Due to its large $N\times N$ size (here, $N\approx 10^5$ is the number of $\mathbf{k}$-points), diagonalizing $\mathbf{A}$ is computationally expensive. In DMD, a key step is rewriting this matrix in the reduced $r$-dimensional space:
\begin{equation}
    \tilde{\mathbf{A}} = \hermconj{\tilde{\mathbf{U}}} \mathbf{A} \tilde{\mathbf{U}} =  
    \hermconj{\tilde{\mathbf{U}}} \mathbf{X}_2 \tilde{\mathbf{V}} \tilde{\mathbf{\Sigma}}^{-1},
\end{equation}
allowing for straightforward eigenvalue decomposition:
\begin{equation}
\label{eq:diago}
    \tilde{\mathbf{A}} \mathbf{W} = \mathbf{W} \mathbf{\Lambda},
\end{equation}
where the matrix $\mathbf{W}$ contains the eigenvectors of $\tilde{\mathbf{A}}$ and the eigenvalues $\mathbf{\Lambda} = \mathrm{diag}\{\lambda_l\}$ are common to both matrices $\tilde{\mathbf{A}}$ and $\mathbf{A}$~\cite{H_Tu_2014}.
The DMD  modes, stacked column-wise in the matrix $\mathbf{\Phi} = \big( \boldsymbol{\phi}^1\; \boldsymbol{\phi}^2 \dotsb \boldsymbol{\phi}^r \big) \in \mathbb{C}^{N\times r}$, can be obtained using~\cite{H_Tu_2014}
\begin{equation}
    \mathbf{\Phi} = \mathbf{X}_2 \tilde{\mathbf{V}} \tilde{\mathbf{\Sigma}}^{-1} \mathbf{W}.
\end{equation}
The DMD frequency of mode $l$ is obtained from the corresponding eigenvalue $\lambda_l$ using equation \ref{eq:diago},

\begin{equation}
    \omega^{\mathrm{DMD}}_l = -i \frac{\ln{\lambda_l}}  {\Delta t},
\end{equation}
where $\Delta t$ is the simulation time step.

The mode amplitudes $\mathbf{b} = \big( b_1\; b_2 \dotsb b_r \big) \in \mathbb{C}^{r}$ are obtained from the initial condition. Setting $t=0$ in equation~\ref{eq:DMD-form}, we get
\begin{equation}
    f_{\mathbf{k}}(0) = \mathbf{\Phi} \mathbf{b},
\end{equation}
and thus the mode amplitude vector $\mathbf{b}$ is obtained from the pseudo-inverse of the DMD mode matrix $\mathbf{\Phi}$:
\begin{equation}
    \mathbf{b} = \pseudoinv{\mathbf{\Phi}} f_{\mathbf{k}}(0).
\end{equation}

This approach provides the DMD modes $\phi^l_\mathbf{k}$, frequencies $\omega_l^{\mathrm{DMD}}$, and mode amplitudes $b_l$, and thus all the quantities needed for DMD prediction of the dynamics outside the sampling window ($t>t_M$) using equation~\ref{eq:DMD-form}. 

\subsection{Electron-phonon scattering from first principles}
\vspace{-10pt}
Our first-principles calculations of $e$-ph scattering employ an established workflow, which is summarized here and described in more detail in Ref.~\cite{zhou2021perturbo}.
The electronic wave functions and band energies are obtained from plane-wave DFT calculations with the {\sc{Quantum Espresso}} code~\cite{QE2017} using the local density approximation~\cite{perdew1992accurate} and norm-conserving pseudopotentials~\cite{troullier1991efficient}. The electronic quasiparticle band structure is refined using GW calculations carried out with the \textsc{YAMBO} code~\cite{sangalliManybody2019}. 
This step improves the agreement with experiment of the electron effective masses and relative valley energies, which are essential for precise calculations of high-field dynamics~\cite{Maliyov_2021} and excited electron relaxation~\cite{Jhalani2017}.
\\
\indent
The phonon dispersion and $e$-ph perturbation potentials are obtained from density-functional perturbation theory (DFPT), where lattice vibrations and their coupling with electrons are treated as perturbations to the ground-state electron density~\cite{QE2017}. The $e$-ph interactions are described by the matrix elements
\begin{equation}
g_{m n \nu}(\mathbf{k},\mathbf{q})=\left(\frac{\hbar}{2 \omega_{\nu \mathbf{q}}}\right)^{1 / 2} 
\left\langle\psi_{m \mathbf{k}+\mathbf{q}} \left| 
\Delta_{\nu \mathbf{q}} V^{\mathrm{KS}} 
\right| \psi_{n \mathbf{k}}\right\rangle,
\end{equation}
which are the probability amplitudes to scatter from an initial electronic state $| \psi_{n \mathbf{k}}\rangle$, with band $n$ and momentum $\mathbf{k}$, to a final state $| \psi_{m \mathbf{k}+\mathbf{q}} \rangle$ by absorbing or emitting a phonon with mode $\nu$, momentum $\mathbf{q}$, and \mbox{frequency $\omega_{\nu \mathbf{q}}$.}
The term $\Delta_{\nu \mathbf{q}} V^{\mathrm{KS}}$ is the $e$-ph perturbation potential induced by the phonon mode and is defined in Ref.~\cite{zhou2021perturbo}.
\\
\indent
To study nonequilibrium dynamics with the rt-BTE, the electrons, phonons, and $e$-ph scattering are described on dense $\mathbf{k}$- and $\mathbf{q}$-point momentum grids. Obtaining the $e$-ph matrix elements on such grids directly from DFPT is computationally prohibitive. Therefore, we first compute $g_{m n \nu}(\mathbf{k},\mathbf{q})$ on coarse $\mathbf{k}$- and $\mathbf{q}$-point grids, and then interpolate these quantities to significantly finer grids using Wannier-Fourier interpolation with Wannier functions generated from {\sc{Wannier90}}~\cite{Mostofi_2014}. 
Finally, the $e$-ph scattering integral employed in equation~\ref{eq:rt-bte} is defined as
\begin{equation}
\label{eq:col-int}
\begin{split}
        \mathcal{I}\big[ f_{n\mathbf{k}} \big] =& -\frac{2\pi}{\hbar} \frac{1}{\mathcal{N}_{\mathbf{q}}} \sum_{m\mathbf{q}\nu} |g_{mn\nu}(\mathbf{k},\mathbf{q})|^2 \\ 
    &\times \Big(\delta (\varepsilon_{n\mathbf{k}}-\hbar\omega_{\nu \mathbf{q}}-\varepsilon_{m\mathbf{k+q}}) \times F_{\mathrm{em}}\big[ f_{n\mathbf{k}} \big] \\
    &+\;\; \delta (\varepsilon_{n\mathbf{k}}+\hbar\omega_{\nu \mathbf{q}}-\varepsilon_{m\mathbf{k+q}}) \times F_{\mathrm{abs}} \big[ f_{n\mathbf{k}} \big] \Big),
\end{split}
\end{equation}
where $\mathcal{N_{\mathbf{q}}}$ is the number of $\mathbf{q}$-points, $\varepsilon_{n\mathbf{k}}$ and  $\varepsilon_{m\mathbf{k+q}}$ are the band energies of the initial and final electronic states, and the Dirac delta functions expressing energy conservation are implemented as Gaussians with a small ($\sim$5~meV) broadening. Above, $F_{\mathrm{abs}}$ and $F_{\mathrm{em}}$ are phonon absorption and emission terms, whose explicit expressions are given in Ref.~\cite{zhou2021perturbo}. 
The Wannier interpolation, scattering integral computation, and the rt-BTE ultrafast dynamics are implemented in our {\sc{Perturbo}} open-source package~\cite{zhou2021perturbo}.

\vspace{6pt}
\begin{acknowledgments}
\vspace{-10pt}
This work was supported by the U.S. Department of Energy, Office of Science, Office of Advanced Scientific Computing Research and Office of Basic Energy Sciences, Scientific Discovery through Advanced Computing (SciDAC) program under Award No. DESC0022088. 
The ultrafast carrier dynamics calculations are based on work performed within the \mbox{Liquid} Sunlight Alliance, which is supported by the U.S. Department of Energy, Office of Science, Office of Basic Energy Sciences, Fuels from Sunlight Hub under Award DE-SC0021266. 
This research used resources of the National Energy Research Scientific Computing Center (NERSC), a U.S. Department of Energy Office of Science User Facility located at Lawrence Berkeley National Laboratory, operated under Contract No. DE-AC02-05CH11231. 

\end{acknowledgments}

\bibliographystyle{mynaturemag-links}
\bibliography{bibliography}

\end{document}